%% file: amr.tex
\documentclass{amsart}[12pt]
\usepackage{epsfig,array,amssymb}

\DeclareFontFamily{OT1}{rsfs}{} \DeclareFontShape{OT1}{rsfs}{m}{n}{
<-7> rsfs5 <7-10> rsfs7 <10-> rsfs10}{}
\DeclareMathAlphabet{\mycal}{OT1}{rsfs}{m}{n}

\def\scri{{\mycal I}}
\def\scrip{\scri^{+}}%

\begin{document}

\markboth{P. Csizmadia}{Testing a new mesh refinement code in the evolution
of a spherically symmetric KG field}

\title{Testing a new mesh refinement code in the evolution
of a spherically symmetric Klein-Gordon field}

\author{P{\'e}ter Csizmadia}


\maketitle

\begin{abstract}
Numerical evolution of the spherically symmetric, massive Klein-Gordon
field is presented using a new adaptive mesh refinement (AMR) code with
fourth order discretization in space and time, along with
compactification in space. The system is non-interacting thus the
initial disturbance is entirely radiated away. The main aim is to
simulate its propagation until it vanishes near $\scrip$. By numerical
investigations of the violation of the energy balance relations, the
space-time boundaries of ``well-behaving'' regions are determined for
different values of the AMR parameters. An important result is that
mesh refinement maintains precision in the central region for
longer time even if the mesh is only refined outside of this region.
The speed of the algorithm was also tested, in case of 10 refinement
levels the algorithm was two orders of magnitude faster than the
extrapolated time of the corresponding unigrid run.
\end{abstract}


\section{Introduction}

The primary motivation to develop the presented techniques and perform the
associated investigations is the need for the efficient simulation of
nonlinear dynamical systems. Numerical integration of nonlinear field
equations is a difficult problem even if the metric is fixed. An obvious
complication is that the field propagates in infinite space-time, but the
computational resources are finite.  Fortunately, infinity can be brought to
finite distance by compactifying space-time, with a conformal rescaling of the
metric \cite{FrauendienerConformalInf}.  A well-chosen coordinate
transformation can also increase the efficiency of numerical calculations.
However, a good transformation is often hard to find, especially when time
evolution changes the system drastically, or in the presence of more than two
different scales.  Some examples are black hole formation, black hole merger,
compact binary stars, etc.  The sizes of such gravitational radiation sources
are very small compared to the produced wavelength, which is much smaller
than the distance from the detector. Different length scales must be
considered simultaneously, but it is extremely hard if time evolution is
simulated on a uniform grid.  Adaptive mesh refinement (AMR) algorithms
overcome these difficulties by using a coarse base grid which is refined
automatically at ``interesting'' locations for more precise calculation
\cite{BergerOliger}.

The precision also depends on the order of numerical schemes used. According
to H{\"u}bner, fourth order calculations are at least two orders of magnitude
more efficient than second order \cite{Hubner1999}. Nevertheless, only second
order calculations are known to be used by AMR codes in numerical relativity
so far, although several implementations of the algorithm were developed since
Choptuik's pioneering work \cite{ChoptuikAMRGravCollapse1993}.  Our choice of
the fourth order Runge-Kutta scheme is also supported by the result of Hansen,
Khokhlov and Novikov. They found that among the methods they investigated,
this is the most efficient one in terms of accuracy and dissipation
\cite{HansenKhokhlovNovikov}.

Problems where gravity is coupled to matters fields are complicated to start
with, thus the code will be applied first to study simpler systems, physical
fields in flat space-time. The simplest possible massive field is the free
Klein-Gordon field, the time evolution of which is investigated in case of
spherical symmetry in this paper. This system is known to provide certain
surprises in numerical simulations \cite{FodorRaczKG}, moreover there is a
straightforward means of checking the efficiency of the developed numerical
method by making use of the analytic solution to the initial value problem.

However, in contrast to the usual scenarios, there is no compact
central object in this system that could keep any part of the initial
disturbance from radiating away. Instead of studying the central
region where vibration with decreasing amplitude is left behind,
the aim is the simulation of the field at larger distances, where
a hypothetical detector could be placed. As the wave propagates
outwards and approaches $\scrip$, its wavelength approaches zero,
partly due to a physical effect \cite{FodorRaczKG}, partly because
of the space compactification. Hence for an accurate simulation of
the long-range behavior, finer and finer subgrids are needed as the
wave gets closer to $\scrip$.

Interaction will be included in forthcoming studies. Plans include the
verification of earlier numerical results on the logarithmic decay of
$\phi^4$ breathers \cite{GeickePhi4} and the study of excitations of
Bogomol'nyi-Prasad-Sommerfield magnetic monopoles.  In the latter
case, a previous study \cite{FodorRaczYMH} will be extended by the
inclusion of the Higgs field's self interaction. In both of these
settings, nonlinear massive fields evolve in fixed Minkowski
space-time.

\section{The AMR algorithm}

The code is based on the Berger-Oliger algorithm \cite{BergerOliger}.  At the
initial time ($T=0$), it starts with a uniform grid, where the values in the
grid points are determined by the initial condition. Based on local error
estimations of the first integration step, child grids may be generated and
filled with values of the initial condition function. This procedure is
applied recursively, until either the local error becomes lower than the
chosen tolerance in each point or the maximum refinement level is reached.
In later integration steps, if a refinement level contains point(s) with
relatively large errors, then finer levels are regridded in child to parent
order (finest first). When a level is regridded, new points are interpolated
using old points from the same level and the parent level.

The same refinement ratio $r$ is used on each level. For the refinement
criteria, Richardson error estimation of an arbitrarily chosen function
$u$ is used. The grid is refined if
\begin{equation}
\frac{|u(Q_{\Delta t,\Delta x}^2 f(x,t)) - u(Q_{2\Delta t,2\Delta x} f(x,t))|}
   {2^{q+1}-2} > \varepsilon\Delta t,
\end{equation}
where $Q_{\Delta t,\Delta x}$ is the numerical integration operator,
$f(x,t+\Delta t) \simeq Q_{\Delta t,\Delta x} f(x,t)$, $q$ is the order of
the method, $f$ is the field and $\varepsilon$ is the error tolerance (see
\cite{BergerOliger}). The exact form of $u$ proved to be irrelevant, provided
that it has an approximate linear dependence on the field components.  For
example, the square root of the energy density was a good choice in
Klein-Gordon simulations, but the value of field $f$ was equally good. For
simplicity, I choose the latter in all simulations.

\begin{figure}[t]
\begin{center}
\epsfig{figure=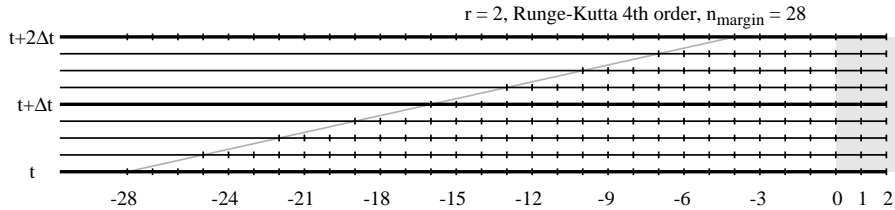,width=12cm}
\end{center}
\caption{Subgrid margin narrowing in case of $r=2$ and 4th order Runge-Kutta.
Margin points -28, ..., -1 at $t$, $t+2\Delta t$, $t+4\Delta t$ etc. are
calculated by interpolation of points in the coarse (parent) grid.}
\label{rk4r2.fig}
\end{figure}

To reach high precision in the numerical calculations, only symmetric finite
difference and interpolation schemes are used whenever it is possible. Space
interpolation is simplified by aligning new subgrids with their parent grid.
Time interpolation can be avoided in one space dimension, because it is
possible to apply relatively large grid margins without noticeable
efficiency loss.  The initial subgrid margin is proportional to $r$ and it
shrinks in each step, until the time becomes equal to the time on the
enclosing coarser grid. (See Fig.~\ref{rk4r2.fig}.) Then the new values on the
margin are determined by space interpolation.

Because of the above-mentioned proportionality, $r$ should not be too large,
otherwise the large margin sizes could result in unnecessary slowdown.  In
case of a larger refinement ratio, less refined levels are needed for the same
precision, but a smaller $r$ has the benefit that the mesh can adapt more
closely to the solution. In my tests, simulations using double refinement were
slightly more efficient (by about 5\%) than triple refinement, thus I choose
$r=2$.

For time and space discretization, the same order of accuracy was
used. Second order Runge-Kutta time integration with second order
space discretization and third order space interpolation, fourth order
Runge-Kutta with fourth order space discretization and fifth order
interpolation.  The presented numerical results are calculated in
fourth order. A fourth order centered finite difference scheme uses
5=2+1+2 points, thus the numerical error propagates 2 points in each
step. However, to avoid instabilities, an artificial dissipation term
containing the sixth order sixth derivative was also applied
~\cite{Hubner1999}:
\begin{equation}
\sigma f'''''' (\Delta x)^5
\ =\ \sigma \cdot (f_{k-3} - 6 f_{k-2} + 15 f_{k-1} - 20 f_k
+ 15 f_{k+1} - 6 f_{k+2} + f_{k+3})/\Delta x,
\end{equation}
where I used $\sigma=0.01$. It increases error propagation velocity to 3
points per step.

Therefore refined grid margins must be shrunk by 3 points in each step. The
fourth order Runge-Kutta method consists of 4 substeps, thus the overall loss
of refined margin points is $12r=24$ in a coarse time step, see
Fig.~\ref{rk4r2.fig}. However, a 24-point wide margin would not be sufficient
if the refined and the coarse grid have their origins (zero index points) or
right edges at the same location.  After a coarse time step, the coarse margin
shrinks to 24-12=12 points. Although the leftmost refined margin point (with
index $i_\mathrm{refined}=-24$) coincides with the leftmost coarse margin
point ($i_\mathrm{coarse}=-12$) at this time, its neighbor
($i_\mathrm{refined}=-23$) is not in a lucky position, it must be interpolated
from coarse margin points. Fifth order centered interpolation requires 3
points on both sides, but there is only one coarse point
($i_\mathrm{coarse}=-12$) on the left side in this case. Choosing a
slightly wider initial margin, with at least 28 points, can solve this
problem. Then the coarse margin shrinks to 28-12=16 points which is
enough to interpolate all refined margin points.

For the fifth order interpolation of field values at point $k$, the following
formula is used:
\begin{equation}
f_k\,\ =\,\ \frac{A_0 f_{i_0} + A_1 f_{i_1} + A_2 f_{i_2} + A_3 f_{i_3}
		+ A_4 f_{i_4} + A_5 f_{i_5}}{\sum_i A_i},
\label{interpol.simple.O5.fk}
\end{equation}
where $i_0$, ..., $i_5$ are the indexes of the six known points (from the
coarse grid and the previous refined grid) and $A_0$, ..., $A_5$ are
coefficients. The 6 coefficients are determined by fixing the function values,
the derivatives and the second derivatives in points $i_2$ and $i_3$.
(Derivatives are approximated in fourth order of accuracy from 2+1+2 points.)
Their values depend on the distance (and existence) of the nearest old fine
grid.
\begin{table}[ph]
\caption{Fifth order interpolation coefficients.}
{\input{o5r2fdinterpol.tbl}
\label{o5interpolation.tbl}}
\end{table}
The possible cases are shown in Table~\ref{o5interpolation.tbl}.

\section{Simulation of a Klein-Gordon field}

A free, spherically symmetric Klein-Gordon field is described by the
equation
\begin{equation}
\partial_t^2\Phi - \frac{1}{r}\partial_r^2(r \Phi) + m^2\Phi\,\ =\,\ 0,
\label{freeKG.eq}
\end{equation}
where $m$ is the mass parameter and Minkowski metric is used,
$ds^2=dt^2-dr^2-r^2 d\sigma^2$.

Fodor and R{\'a}cz \cite{FodorRaczKG} have shown that for arbitrary
initial data with compact support the evolution of this system can
be characterized by the formation of self similarly expanding
shells built up of high frequency oscillations:
\begin{equation}
\Phi(t,r)\,\ \approx
\,\ t^{-3/2}\Psi_{\dot\Phi_0}(r/t)
	\cos\left(m\sqrt{t^2-r^2}+\frac{\pi}{4}\right),
\label{Phi.shells.eq}
\end{equation}
where $\pm\Psi_{\dot\Phi_0}(r/t)$ are the contours of the shells; the
form of this function depends on the initial condition. An important
property of this solution is that both the frequency and the wave
number of oscillations on the outer shell grows proportionally to
$\sqrt{t}$. To prove it, we determine these quantities as the partial
derivatives of the cosine's argument:
\begin{align}
\omega\,\ =&\,\ \frac{\partial(m\sqrt{t^2-r^2})}{\partial t}
\,\ =\,\ \frac{mt}{\sqrt{t^2-r^2}},\\
k\,\ =&\,\ -\frac{\partial(m\sqrt{t^2-r^2})}{\partial r}
\,\ =\,\ \frac{mr}{\sqrt{t^2-r^2}}.
\end{align}
The outer edge expands with the velocity of light, $r\approx t$,
hence the frequency in a short distance $\Delta r=t-r$ from the
outer edge is approximately equal to the wave number, their values
can be estimated using the following formula:
\begin{equation}
\omega\,\ \approx\,\ k
\,\ \approx\,\ \frac{mt}{\sqrt{(\Delta r)(t+r)}}
\,\ \approx\,\ \frac{m\sqrt{t}}{\sqrt{2\Delta r}}.
\label{freq.outershell.eq}
\end{equation}

To remove the $1/r$ coordinate singularity from Eq.~\ref{freeKG.eq}, the
``unphysical'' field
\begin{equation}
f(t,r)\ =\ r\Phi(t,r)
\end{equation}
is used instead of $\Phi$. Space is compactified using a transformation as in
Refs.\cite{FodorRaczYMH,Moncrief}:
\begin{equation}
T(t,r)\ =\ \omega t-\sqrt{\omega^2 r^2+1},\quad
R(r)\ =\ \frac{\sqrt{\omega^2 r^2+1}-1}{\omega r},
\label{coords.TR.eq}
\end{equation}
where $-\infty<T<\infty$ and $0\leq R<1$.

To simplify numerical calculations, the second derivatives are removed from
the field equation by introducing the partial derivatives of $f$ as new
variables $f_T$ and $f_R$. Then the field equation (\ref{freeKG.eq}) can be
written as a system of 3 first order partial differential equations and a
constraint condition:
\begin{align}
\partial_T f\ =&\ f_T,\\
\partial_T f_T\ =&\ - \frac{1-R^2}{1+R^2}f_T
- \frac{R(3+R^2)(1-R^2)}{2(1+R^2)}f_R
- 2R\partial_R f_T + \frac{(1-R^2)^2}{4}\partial_R f_R\nonumber\\
& -\begin{cases}
	\frac{1}{\omega^2}\left(\frac{1+R^2}{1-R^2}\right)^2 m^2 f,
	   & \text{if } R\ne 1,\\
        0, & \text{if } R=1,
   \end{cases}\\
\partial_T f_R\ =&\ \partial_R f_T,\label{dTfT.freeKG.eq}\\
f_R\ =&\ \partial_R f.
\end{align}
{\it Boundary conditions.} For the numerical calculation of derivatives near
$R=0$, information on the parity properties of the functions must be applied.
Whenever the field $\Phi$ is smooth, it has to be an even function of $R$.
Hence $f$ and $f_T$ are odd and $f_R$ is even:
\begin{equation}
f(-R)\ =\ -f(R),\quad f_T(-R)\ =\ -f_T(R),\quad f_R(-R)\ =\ f_R(R).
\end{equation}
The behavior of function $f$ near the other boundary is also important for
similar reasons. The field $\Phi$ vanishes in $\scrip$, thus the function
values for $R\geq 1$ points are supposed to vanish too:
\begin{equation}
f(R)\ =\ 0,\quad f_T(R)\ =\ 0,\quad f_R(R)\ =\ 0.
\end{equation}
The initial condition is a smooth, motionless hunch on the $T(t,r)=0$
hypersurface:
\begin{align}
f\ =&\ \begin{cases}
	c\cdot\exp\left(\frac{d}{(r-a)^2 - b^2}\right)
		& \text{if } r > a - b \text{ and } r < a + b\\
	0 & \text{otherwise},
     \end{cases}\label{f.inithunch.eq}\\
f_T\ =&\ 0\label{fT.inithunch.eq}.
\end{align}
In the simulations, I used a hunch at $R\simeq 0.050\pm 0.037$, with parameters
$\omega=0.05$, $a=2$ (center of hunch in $r$), $b=1.5$ (half-width in $r$),
$c=70$ and $d=10$.

\begin{figure}
\begin{center}
\epsfig{figure=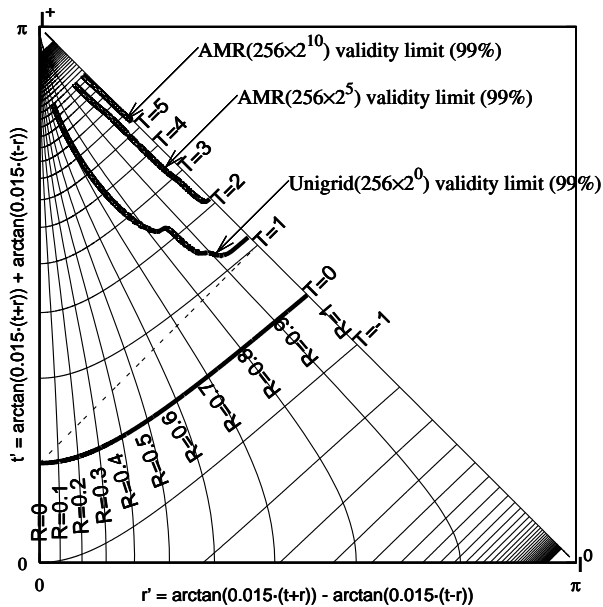,height=8cm}
\end{center}
\caption{Penrose diagram of a part of the Minkowski space-time, with the $R$,
$T$ coordinate lines. Initial condition is specified on the $T=0$ spacelike
hypersurface (thick line). A null geodesic from the origin $(R,T)=(0,0)$
reaches $\scrip$ at $T=1$ (dashed line).  A numerical solution preserves
initial energy with 99\% precision under its ``validity limit'' (upper thick
lines).}
\label{penrose.fig}
\end{figure}
This disturbance is initially close to the origin, thus it is expected to
reach $\scrip$, the future null infinity, at about $T\simeq 1$ like the null
geodesic denoted by dashed line on Fig.~\ref{penrose.fig}.

\section{Results}

\begin{figure}[t]
\begin{center}
\epsfig{figure=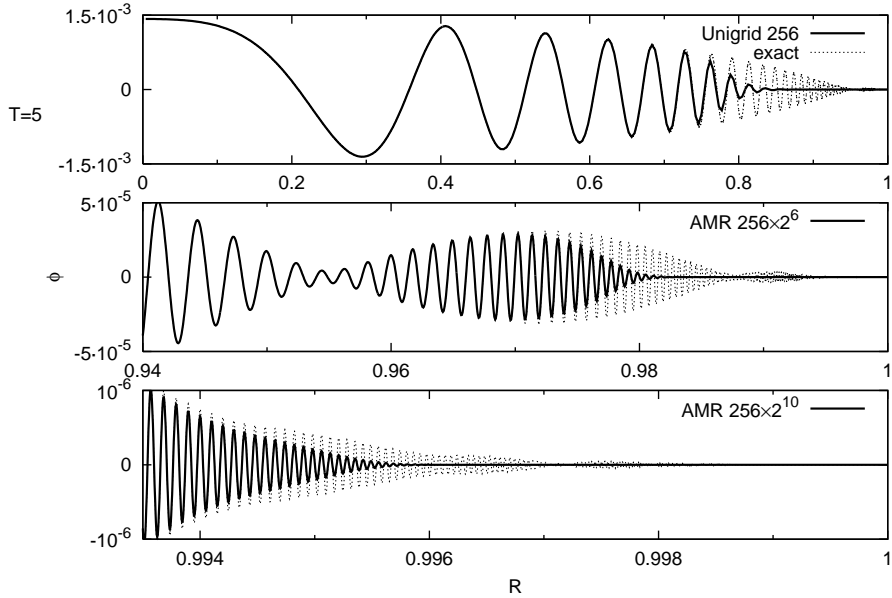,width=12cm}
\end{center}
\caption{Exact (thin dashed lines) and numerically calculated field values
(bold lines) on the $T=5$ hypersurface. The top plot shows the result of a
unigrid simulation with 256 grid points. The right edge is calculated more
precisely, with 6 levels of refinement (middle). An even closer look with 10
levels of refinement is shown on the bottom plot.}
\label{timeslice5.fig}
\end{figure}
The simulations reproduced the most important feature of the
analytical result by Fodor and R{\'a}cz \cite{FodorRaczKG}, the
self similarly expanding shells built up of higher frequency
oscillations. The amplitudes of these high frequency ``carrier
waves'' are modulated by the $\Psi_{\dot\Phi_0}$ function in
Eq.~(\ref{Phi.shells.eq}). As the shells expand and approach
$\scrip$, the {\it $R$-length} (the wavelength in $R$ coordinate
units) of the carrier wave approaches zero, thus finer and finer
grids are needed to simulate its propagation. Accordingly, to zoom
into the vicinity of $R=1$, the number of refinement levels must
be increased, see Fig.~\ref{timeslice5.fig}.

\begin{figure}[t]
\begin{center}
\epsfig{figure=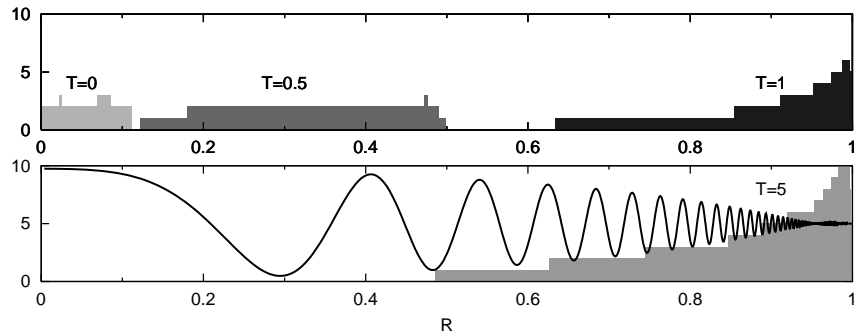,width=12cm}
\end{center}
\caption{Number of mesh refinement levels as a function of $R$ in
different time slices. The top plot shows the refinement in $T=0,
\ 0.5,\ 1$ (gray boxes), the bottom plot shows both the refinement
in $T=5$ (gray boxes) and the form of the $\Phi$ function (black
curve).}
\label{meshref.fig}
\end{figure}
At $T=0$, the grid is only refined in a small range enclosing the
initial hunch. Then the refinement follows the $\Phi$ waves, it moves
outwards and expands. When the waves approach $\scrip$, the peak of
the refinement level function begins to increase, see
Fig.~\ref{meshref.fig}. Since the exact solution of the current
problem is known, it might be possible to replace mesh refinement by a
uniform mesh together with a clever time dependent coordinate
transformation.  The mesh refinement structure shown on
Fig.~\ref{meshref.fig} foreshadows the complications in finding such a
transformation in space; a hunch like this cannot be smoothed using a
simple, monotonous function for $T<1$. There are complications also
for $T>1$, because the frequency of oscillations on the outer shell
increases in time (see Eq.~(\ref{freq.outershell.eq})). Moreover, even
if an easily calculable transformation exists which makes mesh
refinement unnecessary, the equal time surfaces would be different
with the new coordinates, making it hard to transform the results back
to the $(T,R)$ system.

{\it Convergence.} The 4th order convergence of the algorithm was verified by
calculating the
\begin{figure}[t]
\begin{center}
\epsfig{figure=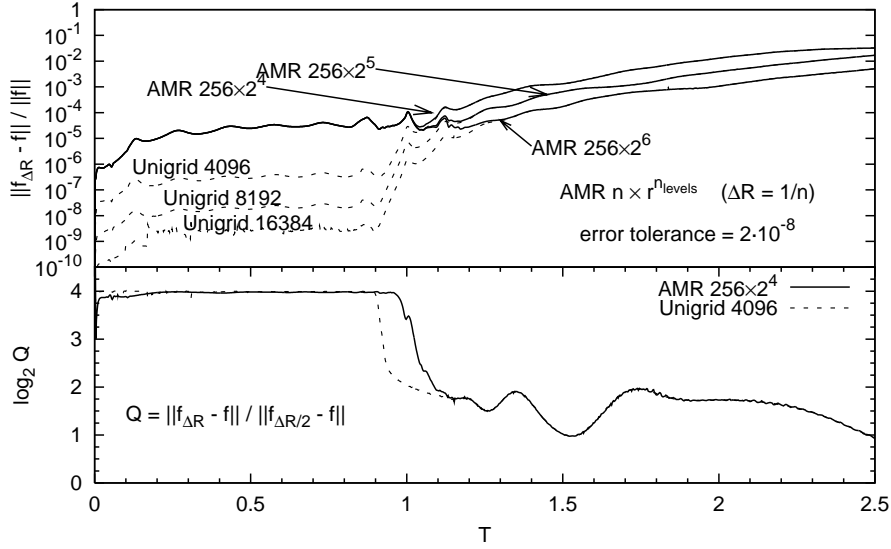,width=12cm}
\end{center}
\caption{Time dependence of the absolute error and the convergence factor in
AMR and unigrid simulations.}
\label{convergence.fig}
\end{figure}
time dependence of the convergence factor
\begin{equation}
Q\ =\ \frac{|| f_{\Delta R} - f ||}
	   {|| f_{\Delta R/2} - f ||},
\end{equation}
where $|| \cdot ||$ is the $L^2$ norm and $f_{\Delta R}$ is the
numerical solution of function $f$ in case of a base grid with $\Delta
R$ spacing.  Locations and sizes of refined subgrids are stored and
reused in calculations with the finer base grids ($\Delta R/2$).
Fig.~\ref{convergence.fig} shows the absolute errors and the
convergence factor for some unigrid and AMR simulations. The $\log_2
Q$ curves start with a plateau at a height of approximately $4$,
proving fourth order convergence.  However, the order of convergence
falls off near $T\simeq 1$ because of the abrupt increase of absolute
error when the wave reaches future null infinity ($\scrip$) and the
$R$-length of oscillations reaches zero, making even the finest mesh
resolution insufficient.

\begin{figure}[t]
\begin{center}
\epsfig{figure=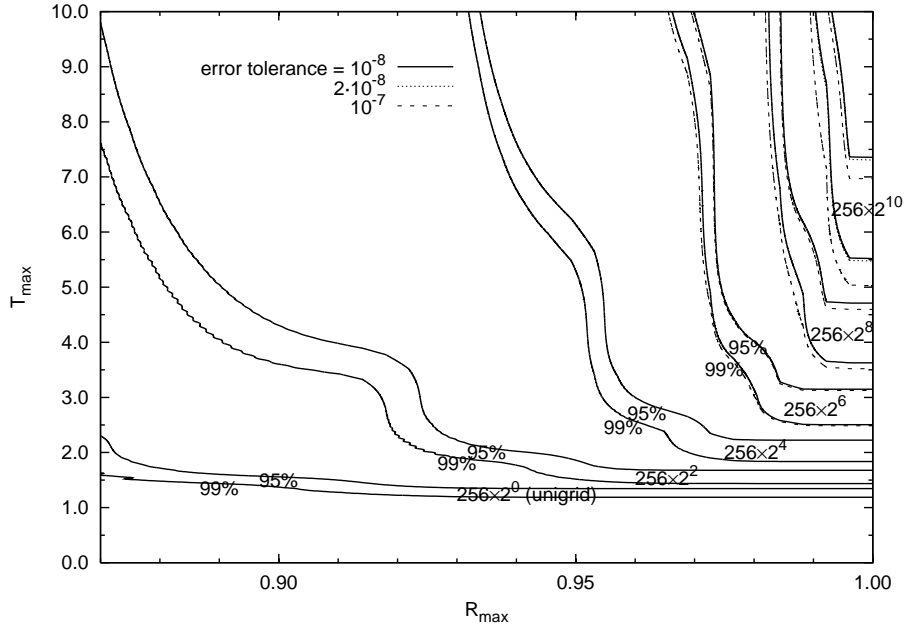,width=12cm}
\end{center}
\caption{Energy conservation bounds of numerical solutions calculated using
different number of refinement levels and different error tolerance settings.}
\label{energy.fig}
\end{figure}
{\it Energy conservation} is violated numerically as the $R$-length of
waves approaching $\scrip$ decrease below grid resolution. To check this, the
following quantity is calculated instead of total energy:
\begin{equation}
E(T_\mathrm{max},R_\mathrm{max}) = \intop_0^{R_\mathrm{max}} dR
	\,\varepsilon(T_\mathrm{max},R)
+ \intop_0^{T_\mathrm{max}} dT\,j_E(T,R_\mathrm{max}),
\end{equation}
where $\varepsilon$ is the energy density and $j_E$ is the energy
flow.  Numerically lost energy is mostly contained in the second term.
By substituting $R_\mathrm{max}=1$, we get the total energy which is
not conserved numerically. To ``restore energy conservation'' at a
given $T$, $R_\mathrm{max}$ must be decreased. Both $T_\mathrm{max}$
and $R_\mathrm{max}$ have a critical value below which energy is
conserved with acceptable precision: $E/E_0 \simeq 1$, where
$E_0=E(0,1)$. Above the critical values, energy is lost: $E/E_0 < 1$.
Fig.~\ref{energy.fig} shows the energy contour lines on the
$T_\mathrm{max}-R_\mathrm{max}$ plane where only 95\% or 99\% of the
initial energy is conserved: $E/E_0=0.95,0.99$.  The conservation
bound can be pushed outwards both in $R_\mathrm{max}$ and
$T_\mathrm{max}$ by increasing the number of refinement levels.

\begin{figure}[t]
\begin{center}
\epsfig{figure=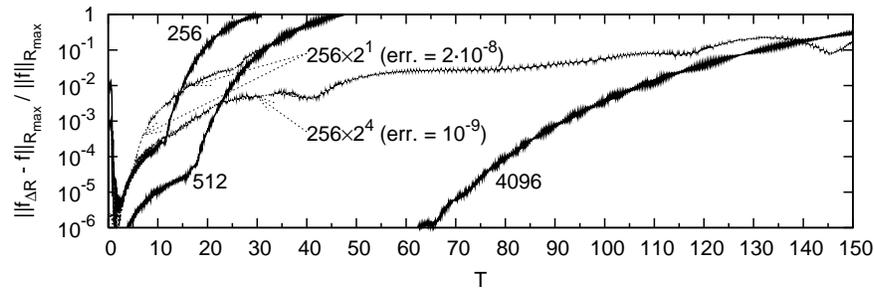,width=12cm}
\end{center}
\caption{Time dependence of the error in the $R \leq R_\mathrm{max} = 0.375$
``central'' range.}
\label{error_near_centre.fig}
\end{figure}
{\it Central range.} As time evolves, the wave packet propagates outwards,
leaving less and less matter in the center. The amplitude of oscillations
decrease asymptotically as $t^{-3/2} \propto T^{-3/2}$ (see Refs.
\cite{FodorRaczKG,HodPiran}). Function $f$ and its derivatives also decrease,
thus one may suppose that a unigrid simulation is enough here. I tested this
assumption by performing unigrid (256, 512 and 4096 points) and AMR
simulations (256 points on base grid, 1 and 4 refinement levels). Errors and
the norm of function $f$ were calculated by restricting integrations to $R
\leq R_\mathrm{max} = 0.375$. Most of the matter leaves this range before
$T\approx 1$, thus function $f$ becomes ``smoother'' and the grid is not
refined at later times here.  Consequently, the error of the $\Delta R=1/256$
unigrid and the same base resolution AMR simulations are close for a while.
However, the unigrid error starts to increase much faster at about $T \approx
12$.  This abrupt degradation proves that error can propagate inwards from
outside ($R > R_\mathrm{max}$). Hence the unigrid error is only ``acceptable''
for $T < 20$, while the 4-level AMR calculation reaches the
same level of inaccuracy much later, at about $T \approx 150$. See the curves
with labels ``$256$'' and ``$256\times 2^4$'' on
Fig.~\ref{error_near_centre.fig}.  On the same plot, it can also be seen that
the error of a high precision unigrid run is smaller than that of the
corresponding AMR with the same maximum precision. Compare the curve labeled
with ``512'' (unigrid) to the ``$256\times 2^1$'' curve (AMR, 1.5 times
faster), and ``$4096$'' to ``$256\times 2^4$'' (AMR, 4 times faster).
However, the AMR error curve increases much slower than the unigrid error
curve, hence they meet at a certain time after which AMR is more accurate.
If the AMR error tolerance is small enough, then the error at their meeting
point is also small.

Note that by restricting attention to the central range in $R$, a much
longer simulation was possible in $T$. This feature can also be seen
on the energy conservation curves on Fig.~\ref{energy.fig}, where
$T_\mathrm{max}$ increases when $R_\mathrm{max}$ is decreased.

\begin{table}
\caption{Speed comparison of unigrid and AMR runs from $T=0$ to $5$. The
``$n$'' columns contain the number of points on the (spacelike) base grid.
Run times ($t_0$ and $t_\mathrm{AMR}$) were measured on AMD Opteron (64 bit)
1.8 GHz hardware and Sun Java 1.5.0\_02 virtual machine.  $N_0$ and
$N_\mathrm{AMR}$ are the total number of points of the spacetime grid.
$^*$Instead of performing unigrid runs with 32768 and more grid points, their
times were extrapolated using the time of the $n=16384$ run and assuming
$t_\mathrm{unigrid}\propto n^2$.}
{\input{speed.tbl}
\label{speed.tbl}}
\end{table}
{\it Speed tests.} The time of a unigrid simulation is proportional to the
total number of grid points in the spacetime domain:
\begin{equation}
t_\mathrm{run}\propto 1/(\Delta x)^{d},\label{runtimefit.eq}
\end{equation}
where $\Delta x\propto 1/n$ is the grid spacing and $d$ is the number of
dimensions of the spacetime grid, $d=2$ in this case. When using AMR, the same
resolution can be reached much faster because only a small part of the grid is
refined. In case of 10 refinement levels, AMR would be two orders of magnitude
faster than a corresponding unigrid run. AMR/unigrid run time ratios can be
approximated by calculating the ratio of the total number of grid points in
the spacetime domain ($N_0$ for unigrid and $N_\mathrm{AMR}$ for AMR), then
adding the overhead ($\sim 30\%$) of the AMR algorithm. The
$t_\mathrm{AMR}/t_0 \sim 1.3 N_\mathrm{AMR}/N_0$ formula is a good
approximation for the measured times in Table~\ref{speed.tbl}, in case of 5 or
more refinement levels. The formula (\ref{runtimefit.eq}) can be fitted nicely
to AMR run times also. These fittings were performed at constant error
tolerance values ($10^{-8}$, $2\cdot 10^{-8}$ and $10^{-7}$) for simplicity.
The result is that for small $T_\mathrm{max}$ values, the ``effective
dimension'' $d$ increases with $T_\mathrm{max}$ until a plateau is reached
near $T_\mathrm{max}=5$, at a height of $d=1.35\pm 0.04$.

\section{Other tests}

Further testing was performed with nonlinear 1 dimensional
Klein-Gordon fields and periodic boundary conditions,
$\Phi(x\pm 1)=\Phi(x)$, by adding a fourth order self interaction
term to the Lagrangian:
\begin{equation}
L\,\ =\,\ {1\over 2}(\partial_t\Phi)^2 - {1\over 2}(\partial_x\Phi)^2
	- \frac{m^2}{2}\Phi^2 - {c_4\over 4}\Phi^4.
\end{equation}
The initial condition at $t=0$ is the motionless hunch used earlier
in Eqs. (\ref{f.inithunch.eq}-\ref{fT.inithunch.eq}) but substituting
$r$ by $x$, $f$ by $\Phi$, $f_T$ by $\Phi_t$ and using the parameter
values $a=0.5$, $b=0.2$, $c=10$, $d=1$. Test problems include
massive ($m^2=1$, $c_4=\frac{1}{3}$ and $m^2=10$,
$c_4=\frac{10}{3}$), massless ($m^2=0$), free ($c_4=0$) cases and the
sine-Gordon field:
\begin{equation}
L\,\ =\,\ {1\over 2}(\partial_t\Phi)^2 - {1\over 2}(\partial_x\Phi)^2
	+ m^2(\cos\Phi - 1).
\end{equation}
In each of these cases, the initial hunch at $x=0.5$ splits and its
two parts propagate in opposite directions. At $t=0.5$ they reach the
periodic boundaries $x=0,\,1$ where they pass through each other.
They meet again at $(t,x)=(1,\ 0.5)$, almost restoring the initial
state but in a distorted form. As a consequence of the translation
invariance and the lack of nonlinear coordinate transformations, the
derivatives and the numerical error do not increase substantially
during the movement of the hunches, the number of required mesh
refinement levels for a given error tolerance is determined by the
initial state and preserved later.

Another test was performed with the $\phi^4$ breather described
by the Lagrangian:
\begin{equation}
L(\phi,\partial_t\phi,\partial_x\phi)\,\ =\,\ \frac{1}{2}(\partial_t\phi)^2
 - \frac{1}{2}(\partial_x\phi)^2
 - \frac{1}{4}(\phi^2-1)^2.
\end{equation}
For the numerical simulation of this system, I compactified space
using the same transformation as earlier in
Eq.~(\ref{coords.TR.eq}), $R=R(x)$ with $\omega=0.05$. Time $t$ is
not transformed. The initial condition is the same as that used by
Geicke \cite{GeickePhi4}, it contains a kink and an antikink in
$x=\pm 0.8$. The radiation of the initial ``kick'' involves the
outwards propagation of a wave with decreasing $R$-length. As it
propagates outwards, more and more levels of refinement are
activated until it vanishes (in $t\gtrsim 10^3$ if the base grid
spacing is $\Delta R\leq 1/128$). Then only the smooth, stable,
oscillating wave remains near the origin and refinement is no
longer needed. I performed simulations without further energy
loss until $t=10^7$ using a $\Delta R=1/128$ base grid. However,
I found energy loss in case of finer base grids. I also found that
increased precision in the simulation of the initial radiation does
not necessarily increase precision near the origin in the long run.
The same amount of energy disappears with the same rate if mesh
refinement is not used at all in case of $\Delta R=1/384$ and
$t\leq 7000$. Further experimenting is needed for the
verification of the statement of Geicke \cite{GeickePhi4} about the
logarithmic decay of this system.

\section{Summary}

A new AMR code was developed for integrating field equations
numerically in time. It is tested thoroughly using fourth order
Runge-Kutta method and fourth order space discretization, but it is
also possible to use other numerical schemes. The main test problem is
the simulation of a spherically symmetric Klein-Gordon field in a
special coordinate system (\ref{coords.TR.eq}) with compactified space
coordinate. The exact solution of this problem is known, thus it is
possible to calculate the absolute error of the numerical simulation.
By calculating the errors of AMR simulations with different base grid
spacings, the fourth order convergence of the algorithm was shown.
The numerical violation of energy conservation was also investigated;
I determined the space-time boundaries of the ``well-behaving'' range.
Inside this space-time volume, the sum of the total energy and its
integrated outgoing flux is constant with an acceptable precision. The
time boundary $T_\mathrm{max}$ is a monotonically decreasing function
of the maximum space coordinate $R_\mathrm{max}$ (see
Fig.~\ref{energy.fig}).  It means that energy is conserved numerically
for longer time if a smaller central range is examined in space. The
error of the simulation behaves similarly; longer runs can be
``closer'' to the exact solution if the error norm is calculated only
in a small central range. The speed of the algorithm was also tested,
in case of 10 refinement levels the AMR algorithm was two orders of
magnitude faster than the extrapolated time of the corresponding
unigrid run.

\section*{Acknowledgments}

I would like to thank Istv{\'a}n R{\'a}cz and Gyula Fodor for the suggestions,
especially for the initial idea to develop an AMR code. I also thank them for
the stimulating discussions.

\end{document}

%% file: o5r2fdinterpol.tbl
\setlength{\tabcolsep}{2pt}\begin{tabular}{
>{\small\ttfamily}c>{\small\ttfamily}c>{\small\ttfamily}c>{\small\ttfamily}c
>{\small\ttfamily}c>{\small\ttfamily}c>{\small\ttfamily}c>{\small\ttfamily}c
>{\small\ttfamily}c>{\small\ttfamily}c>{\small\ttfamily}c
|rrrrrr|c}
\,0\, & \,1\, & \,2\, & \,3\, & \,4\, & \,5\, & \,6\, & \,7\, & \,8\, & \,9\,
& 10
 & $A_0$ & $A_1$ & $A_2$ & $A_3$ & $A_4$ & $A_5$ & $\sum_i A_i$\cr
\hline
$i_0$ & & $i_1$ & & $i_2$ & $k$ & $i_3$ & & $i_4$ & & $i_5$
& 3 & -25 & 150 & 150 & -25 & 3 & 256\cr
$i_0$ & & $i_1$ & & $i_2$ & $k$ & $i_3$ & $i_4$ & $i_5$ & $\cdot$ & $\cdot$
& 15 & -116 & 690 & 1020 & -384 & 55 & 1280\cr
$i_0$ & $i_1$ & $i_2$ & $k$ & $i_3$ & & $i_4$ & & $i_5$ & &
& 55 & -384 & 1020 & 690 & -116 & 15 & 1280\cr
\end{tabular}

%% file: speed.tbl
\setlength{\tabcolsep}{2pt}\begin{tabular}{r|r||l|rrr|rr}
\multicolumn{2}{c||}{unigrid} & \multicolumn{6}{c}{AMR}\cr
\hline
& & \multicolumn{1}{c|}{} & \multicolumn{3}{c|}{error tolerance}
& \multicolumn{1}{c}{} & \multicolumn{1}{c}{}\cr
& \multicolumn{1}{c||}{} & & \multicolumn{1}{c}{$10^{-8}$}
& \multicolumn{1}{c}{$2\cdot 10^{-8}$}
& \multicolumn{1}{c|}{$10^{-7}$} & &\cr
\multicolumn{1}{c|}{$n$} & \multicolumn{1}{c||}{$t_0$}
& \multicolumn{1}{c|}{$n\times r^{levels}$}
& \multicolumn{1}{c}{$t_\mathrm{AMR'}$}
& \multicolumn{1}{c}{$t_\mathrm{AMR}$}
& \multicolumn{1}{c|}{$t_\mathrm{AMR''}$}
& \multicolumn{1}{c}{$N_0/N_\mathrm{AMR}$}
& \multicolumn{1}{c}{$t_0/t_\mathrm{AMR}$}\cr
\hline
   256 &   5.9s & & & & & &\cr
   512 &  22.5s & $256\times 2^1$ & 18.8s & 17.6s & 17.3s & 1.7 & 1.3\cr
  1024 &    73s & $256\times 2^2$ & 56s & 46s & 38s & 2.7 & 1.6\cr
  2048 &   295s & $256\times 2^3$ & 118s & 108s & 84s & 4.5 & 2.7\cr
  4096 &  1407s & $256\times 2^4$ & 293s & 263s & 194s & 7.7 & 5.3\cr
  8192 &  5597s & $256\times 2^5$ & 630s & 586s & 468s & 12.9 & 9.6\cr
 16384 & 22360s & $256\times 2^6$ & 1577s & 1416s & 1164s & 20.9 & 15.8\cr
 32768 & 89440s$^*$ & $256\times 2^7$ & 4110s & 3574s & 2916s & 32 & 25\cr
 65536 & 357760s$^*$ & $256\times 2^8$ & 10904s & 9590s & 8202s & 48 & 37\cr
131072 & 1431040s$^*$ & $256\times 2^9$ & 29064s & 26614s & 21427s & 72 & 54\cr
262144 & 5724160s$^*$ & $256\times 2^{10}$ & 70148s & 63562s & 47825s & 116 & 90\cr
\end{tabular}